\documentclass[letterpaper, 10 pt, conference]{ieeeconf}  

\IEEEoverridecommandlockouts                              
\usepackage[utf8]{inputenc}
\usepackage[T1]{fontenc}

\usepackage{amsmath,amssymb,amsfonts}
\usepackage{bm,bbm,siunitx,hyperref,graphicx}
\usepackage{float}	
\usepackage{xcolor}

\let\labelindent\relax

\usepackage{enumitem}
\usepackage{cite}
\usepackage{balance}

\title{\LARGE \bf
Adaptive Feedforward Reference Design for Active Vibration Rejection in Multi-Actuator Hard Disk Drives
}

\author{Zhi Chen$^{1}$ Nikhil Potu Surya Prakash$^{2}$ and Roberto Horowitz$^{3}$ 
\thanks{$^{1}$Zhi Chen is with Department of Mechanical Engineering,
        University of California, Berkeley, CA 94720, USA
        {\tt\small chenzhi@berkeley.edu}}%
\thanks{$^{2}$Nikhil Potu Surya Prakash is with Department of Mechanical Engineering,
        University of California, Berkeley, CA 94720, USA
        {\tt\small nikhilps@berkeley.edu}}%
\thanks{$^{3}$Roberto Horowitz is with Department of Mechanical Engineering,
        University of California, Berkeley, CA 94720, USA
        {\tt\small horowitz@berkeley.edu}}%
        
}

\begin{document}

\maketitle
\thispagestyle{empty}
\pagestyle{empty}

\begin{abstract}

In December 2017, Seagate unveiled the Multi Actuator Technology to double the data performance of the future generation hard disk drives (HDD). This technology will equip drives with two dual stage actuators (DSA) each comprising of a voice coil motor (VCM) actuator and a piezoelectric micro actuator (MA) operating on the same pivot point. Each DSA is responsible for controlling half of the drive’s arms. As both the DSAs operate independently on the same pivot timber, the control forces and torques generated by one can affect the operation of the other and thereby worsening the performance drastically. In this paper, a robust adaptive feedforward controller is designed as an add-on controller to an existing stabilizing feedback controller to reject the disturbances transferred through the common pivot timber by shaping the references to the Voice Coil Motor (VCM) actuator and the total output of the dual stage system.

\end{abstract}

\section{Introduction}\label{sec:intro}
In the recent years, HDDs have been replaced by solid state drives (SSD) in most of personal
computers regardless of their high cost. One of the most important reasons is that the data
transfer rate of an SSD is much faster than that of an HDD. In 2017, Seagate unveiled its new
multi actuator technology as a breakthrough that can double the data transfer performance
of the future-generation hard drives for hyper-scale data centers. In a multi actuator drive,
the read-write heads are split into two sets, an upper and a lower half, which can double
the data transfer rate by having the upper and lower platter sets work in parallel.

The multi actuator setup brought some new challenges to the HDD's controller design.
Since both the DSAs operate on the same pivot point, the forces and torques generated by
one actuator can affect the operation of the other DSA. The interaction of
the two DSAs can be categorized into three basic scenarios. In the first scenario, both
DSAs are in the track following mode, and it is expected that the interaction between the
two actuators is negligible. In the second scenario, both the DSAs are in seek mode. In
this mode, the coupling vibrational interaction is usually negligible compared to the large
trajectories for both the DSAs. In the third scenario, one DSA is in seek mode
and the other DSA is in track following mode. Under this scenario, the seeking DSA
will impart disturbances in the form of vibration to the track following DSA, which
drastically hamper the performance of the track following DSA.

A data-driven control design approach \cite{c16,c17} to suppress the vibrations generated by the seeking DSA has been applied to a multi actuator drive. The data-driven controllers are robust to the variations in the plant models. However, a common controller might not be the optimal controller
for each individual HDD as there might be variations in the plants. To address this issue, an adaptive
feedforward reference design technique is developed in this work. This work utilizes the reference design for track seeking controllers developed in \cite{c11} to shape the total output of the dual stage system. Since the track seeking architecture is already built inside the HDD, adding the current adaptive control strategy to the HDD would be simple. A pretraining strategy to prevent the MA's output from exceeding its stroke limits during the adaptation process is also presented. 
\section{Control Architectures for Dual Stage Actuators}\label{sec:CtrlArch}
\subsection{Feedback Control Structure}
\begin{figure}[thpb]
    \centering
    \includegraphics[width=\columnwidth]{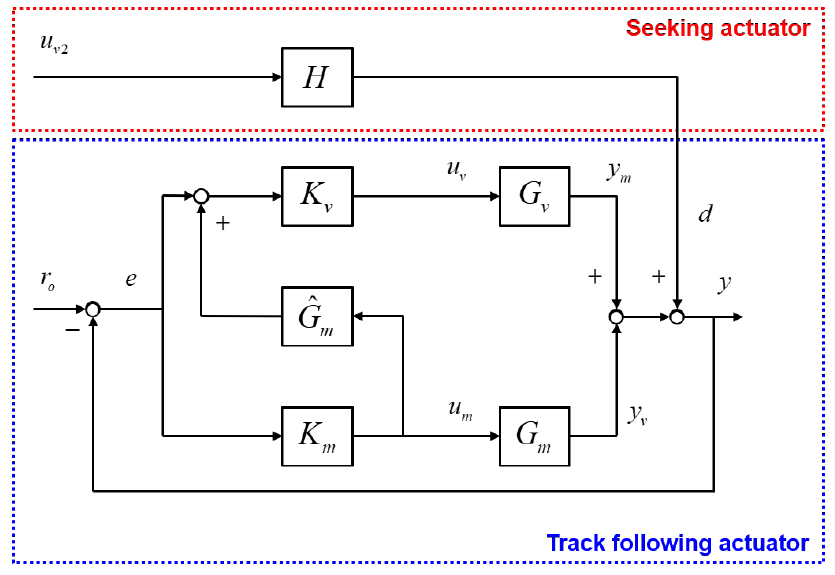}
    \caption{Block diagram of the multi actuator HDD. One of the DSA is in seek mode
and the other is working in the track following mode.}
    \label{fig:TrackFollowingStructure}
\end{figure}
Fig.\ref{fig:TrackFollowingStructure} shows a block diagram of the dual-stage controller
design using sensitivity decoupling method \cite{c9}. $G_{v}$, $G_{m}$ represent the actual plants of the VCM and MA respectively. $K_{v}$ and $K_{m}$ are the Single-Input Single-Output (SISO) controllers for the VCM and MA plants respectively. $\hat{G}_{m}$ is the feedforward plant estimate of MA. $H$ represents the disturbance cross transfer function from the seeking DSA to the track following DSA which takes in the input of the VCM of the track seeking DSA ($u_{v2}$) and produces the disturbance $d$ on the output of the track following DSA. The signals $r_o$ represents the runout, $e$ represents the position error signal (PES), $u_v$ represents the control input to the VCM, $u_m$ represents the control input to the MA, $y_v$ represents the output of VCM, $y_m$ represents the output of MA, $y_t = y_v+y_m$ represents the total output produced by both the actuators and $y$ represents the actual output of the dual stage system along with the disturbance.
The overall sensitivity transfer function from $r_o$ to $e$ can be calculated as 
 \begin{align}\label{eq:sensitivity}
     S &= \frac{1}{1+G_vK_v+G_mK_m+G_vK_v\hat{G}_mK_m} \\
     &= \frac{1}{1+K_vG_v}.\frac{1}{1+K_m\Bar{G}_m}
 \end{align}
 where the modified MA plant $\Bar{G}_m$ is given by
 \begin{equation}
     \Bar{G}_m = G_m +\frac{K_vG_v(\hat{G}_m-G_m)}{1+K_vG_v}
 \end{equation}
 The MA has high frequency uncertainties and hence the estimated model of the MA ($\hat{G}_m$) is a good approximation of the actual MA plant ($G_m$) in the low frequency region. As a
result $\Bar{G}_m$ will be very close to $G_m$ in those low frequency regions. In the high frequency
regions, the term $K_vG_v$ is relatively small as the VCM actuator is not active in that region.
Consequently, any difference between $G_m$ and $\Bar{G}_m$ at high frequency regions will be
decreased by the $K_vG_v$ term and hence $\Bar{G}_m$ will also be a good approximation of $G_m$
at high frequency regions. Therefore, the overall sensitivity of the dual-stage actuation in \ref{eq:sensitivity} can be approximately decoupled as the multiplication of the actuators' sensitivity
transfer functions
\begin{equation}
    S\approx S_vS_m
\end{equation}
where
\begin{equation}
    S_v = \frac{1}{1+K_vG_v}
\end{equation}
\begin{equation}
    S_m = \frac{1}{1+K_mG_m} 
\end{equation}

\subsection{Feedforward Reference Design Structure}
\begin{figure}[thpb]
    \centering
    \includegraphics[width=\columnwidth]{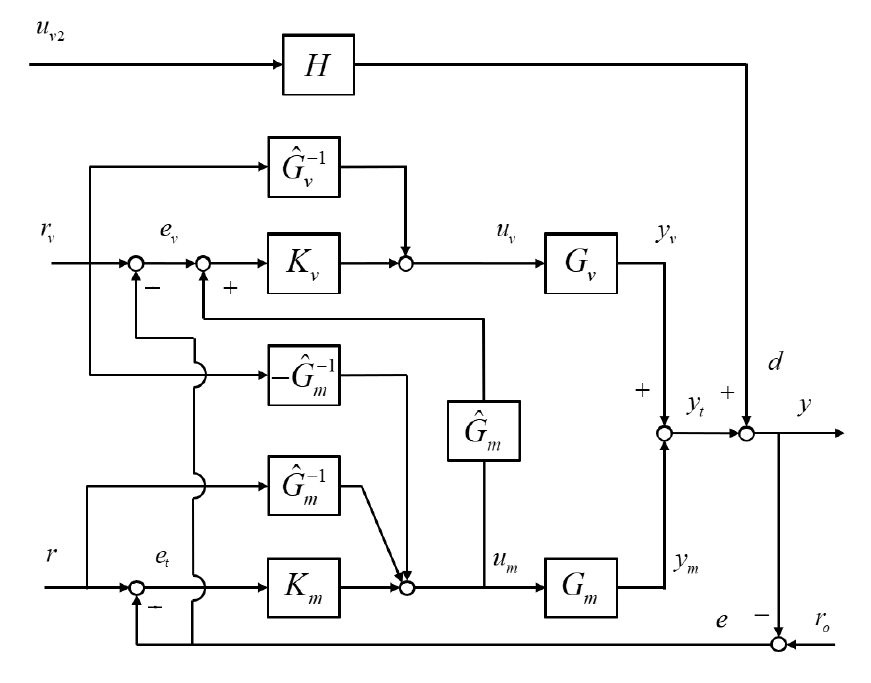}
    \caption{Block diagram of the decoupled track seeking control structure for a dual-stage
servo system.}
    \label{fig:TrackSeekingStructure}
\end{figure}

The feedback control structure for track following shown in the previous section has been extended to track seeking in \cite{c11}. In this structure, the reference trajectories to be followed by the read-write head of the DSA and the VCM actuator are designed to minimize the output of the MA as it has small stroke limits. Fig. \ref{fig:TrackSeekingStructure} shows a block diagram of the feedforward reference design structure. In this structure, $\hat{G}_v$ represents the estimate of the VCM plant. $\hat{G}_v^{-1}$ and $\hat{G}_m^{-1}$ are the Zero Phase plant inverses used in Zero Phase Error Tracking Control (ZPETC) \cite{c21}. The signals $r_v$ represents the reference signal to VCM and $r$ represents the reference to the total output of the dual stage system.  The signals $e_v = r_v-e$ and $e_t = r-e$. 
Various signals and overall transfer functions can be easily obtained by simplifying the block diagram. Let $r_{os}=Sr_o$ and $d_s = Sd$ be the signals obtained by filtering the runout and disturbance signals through the sensitivity transfer function respectively. The PES in the presence of disturbance can be obtained from the block diagram as
\begin{equation}
    e = r_{os}-d_s+R_{r_v\rightarrow e}r_v+R_{r\rightarrow e}r
\end{equation}
where
\begin{equation}\label{eq:rrvtoe}
    R_{r_v\rightarrow e} = S(G_m\hat{G}_m^{-1}-G_v\hat{G}_v^{-1})
\end{equation}
\begin{equation}\label{eq:rrtoe}
    R_{r\rightarrow e} = -S(G_m\hat{G}_m^{-1}+G_vK_v+G_mK_m+G_vK_v\hat{G}_mK_m)
\end{equation}
Here $R_{r_v \rightarrow e}$ denotes the transfer function from $r_v$ to $e$ and $R_{r \rightarrow e}$ denotes the transfer function from $r$ to $e$. If $G_v \hat{G}_v^{-1} \approx 1$ and $G_m \hat{G}_m^{-1} \approx 1$, then $R_{r_v \rightarrow e}$ is close to zero and $R_{r \rightarrow e}$ is close to negative unity.
\begin{equation}
    R_{r_v \rightarrow e} \approx S(1-1) = 0
\end{equation}
From \eqref{eq:sensitivity} and \eqref{eq:rrtoe}
\begin{align}
    R_{r \rightarrow e} &\approx -S(1+G_vK_v+G_mK_m+G_vK_v\hat{G}_mK_m) \nonumber \\&= -1
\end{align}

Furthermore, the PES can be approximated by
\begin{equation}
    e \approx -d_s-r+r_{os}
\end{equation}
It can be seen that if the total reference signal $r$ is designed to be equal to $-d_s$, the PES can be minimized. To achieve this, the disturbance $d_s$ needs to be known. This motivates the design of an adaptive controller to estimate this signal.


\section{Adaptive Control Structure}\label{sec:AC}
\begin{figure}[thpb]
    \centering
    \includegraphics[width=\columnwidth]{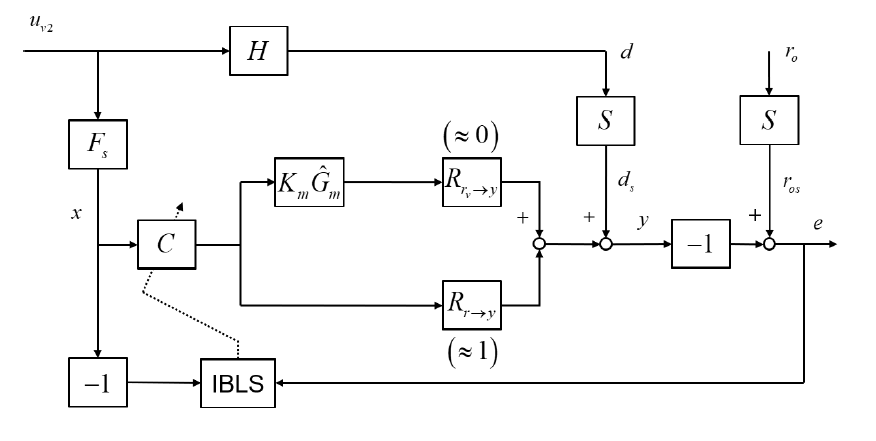}
    \caption{Simplified block diagram of an adaptive feedforward control system.}
    \label{fig:SimplifiedBDofACS}
\end{figure}

In this section, an adaptive controller '$C$' will be designed to estimate the signal $d_s$ and compensate for it. This design utilizes the feedforward reference design structure presented in the previous section. While estimating $d_s$, it is also essential to respect the MA stroke limits and hence the references will be chosen in such a way that the MA's output is minimized.
The output of the MA $y_m$ can be obtained by simplifying the block diagram in fig. \ref{fig:TrackSeekingStructure} as 
\begin{multline}
    y_m  = -G_m\hat{G}_m^{-1}r_v+(G_m\hat{G}_m^{-1}+G_mK_m)r \\
    +G_mK_m(r_{os}-d_s+R_{r_v \rightarrow e}r_v+R_{r\rightarrow e}r)
\end{multline}
Again, if $G_v \hat{G}_v^{-1} \approx 1$ and $G_m \hat{G}_m^{-1} \approx 1$, the following approximation can be obtained for $y_m$. 
\begin{align}\label{eq:ym}
    y_m &\approx -r_v+(1+G_mK_m)r+G_mK_m(r_{os}-d_s-r) \nonumber \\ 
    &= -r_v+r+G_mK_m(-d_s+r_{os})
\end{align}
With the optimal filter $C^*$, the reference signal follows the negative value of the filtered disturbance.
\begin{equation}\label{eq:rapproxds}
    r \approx -d_s
\end{equation}
By substituting \eqref{eq:rapproxds} in \eqref{eq:ym}, and by designing $r_v$ as
\begin{equation}
    r_v = \hat{G}_mK_mr
\end{equation}
we can obtain $y_m$ as

\begin{align}
    y_m &\approx -\hat{G}_mK_mr+r+G_mK_m(-d_s+r_{os}) \nonumber \\
    &\approx \hat{G}_mK_md_s-d_s+G_mK_m(-d_s+r_{os}) \nonumber \\
    &\approx -d_s+G_mK_mr_{os}
\end{align}
As the quantity $G_mK_mr_{os}$ is very small, the MA will just have to track $-d_s$ which is well within its stroke limits.
Similarly, we can also obtain the output of VCM, $y_v$ as
\begin{align}
    y_v &= r_o-e-d-y_m \nonumber \\
    &\approx r_o-(-d_s-r-r_{os})-d-(-d_s+G_mK_mr_{os}) \nonumber \\
    &\approx r_o - r_{os}-d+d_s-G_mK_mr_{os}
\end{align}
It can be seen that the VCM takes care of rejecting most of the actual disturbance as the signal $d$ explicitly appears in the expression for $y_v$.
The total output of the dual stage system $y_t = y_v+y_m$ can be obtained as
\begin{equation}
    y_t = -d+r_o-r_{os}
\end{equation}
The actual output of the dual stage system along with the disturbance would now be
\begin{equation}
    y = y_t+d = r_o-r_{os} \approx 0
\end{equation}
\begin{figure}[thpb]
    \centering
    \includegraphics[width=\columnwidth]{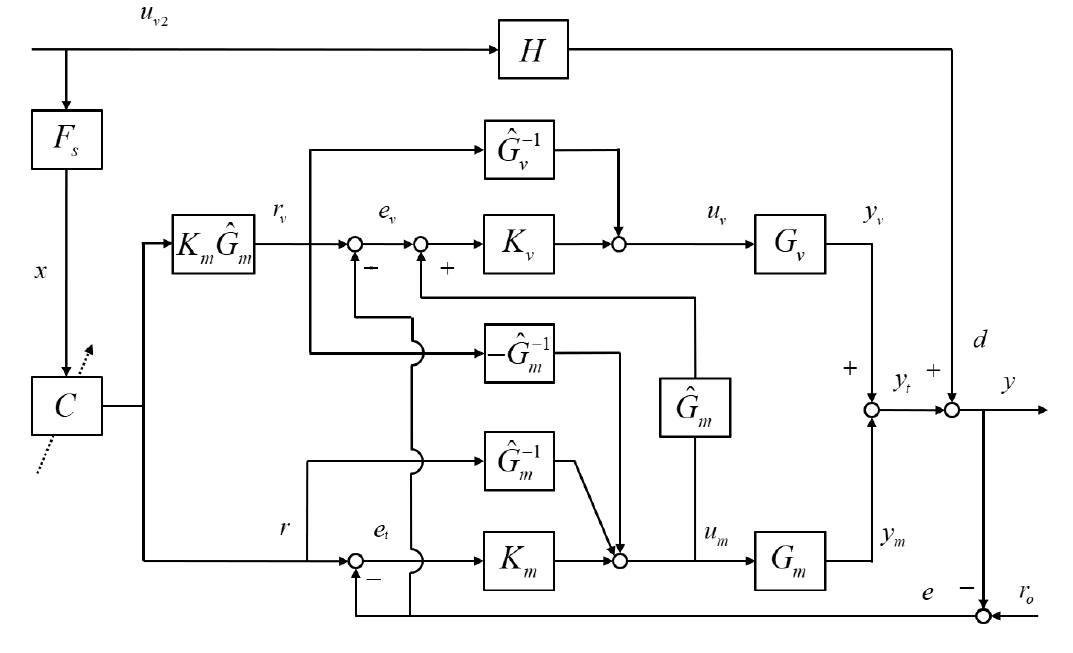}
    \caption{Block diagram of the adaptive feedforward control system.}
    \label{fig:AdaptiveControlStructure}
\end{figure}
Having designed $r_v$, the only task remaining is to make the reference signal $r$ as close to $-d_s$ as possible. This is achieved using the control structure in fig.\ref{fig:AdaptiveControlStructure}. '$C$' represents the adaptive controller and the filter $F_s$ is transfer function containing any known transfer functions. The signal $x$ is $u_{v2}$ filtered through $F_s$. We use an FIR filter with unknown coefficients to design $C$. Any of the available parameter adaptation algorithms can be used to update the coefficients of $C$. In our simulations, we used an Iterative Batch Least Squares (IBLS) approach developed in \cite{c22} to update the parameters of $C$. This approach collects the signals $x$ and $e$ and formulates a least squares optimization problem which is then solved by a stochastic newton conjugate gradient method.
The block diagram shown in fig.\ref{fig:AdaptiveControlStructure} can be simplified to the block diagram shown in fig.\ref{fig:SimplifiedBDofACS}. From fig.\ref{fig:SimplifiedBDofACS}, we can obtain the overall transfer function of the secondary path $P_{dual}$ as  
\begin{equation}\label{eq:pdual}
    P_{dual} = -(R_{r\rightarrow y}+R_{r_v\rightarrow y}\hat{G}_mK_m) \approx -1
\end{equation}
Therefore we can define an estimate of $P_{dual}$ as $\hat{P}_{dual} = -1$ to be used in the parameter adaptation algorithm. The convergence condition for the parameters in the adaptive controller $C$ is stated in the following corollary.\\
\textbf{Corollary 1:} \textit{The controller $C$ converges to a unique stationary point $\hat{C}$ if and only if $P_{dual}/\hat{P}_{dual} = R_{r\rightarrow y}+R_{r_v \rightarrow y}\hat{G}_mK_m$ is strictly positive real (SPR).}

\subsection*{Controller Pretraining on the Single-Stage
Actuator}
As the MA has a stroke limit of only a few micrometers, the adaptive reference
design should take into consideration the saturation properties of the PZT actuator during
the entire parameter adaptation process. As discussed in the previous section, the output of
the MA follows the opposite residual disturbance ($-d_s$) and its absolute value remains
small if the controller parameters are close to the stationary point. If the controller parameters
are far from the optimal, the plant model of the MA becomes nonlinear and
the parameter adaptation algorithm may diverge.

In order to get a good initial point for the adaptive controller in the dual-stage actuator
system, the controller is pretrained on the single-stage actuator system with the VCM. The entire parameter adaptation process is then separated into two stages: a pretraining stage
and a fine-tuning stage. In the pretraining stage, the MA is turned off. The
controller parameters are initialized with some random points and updated on the single-stage
actuator system until they converge. In the fine-tuning stage, the MA is
turned on again. The controller parameters are tuned for the dual-stage actuator system on
the most resent measurements.
\begin{figure}[thpb]
    \centering
    \includegraphics[width=\columnwidth]{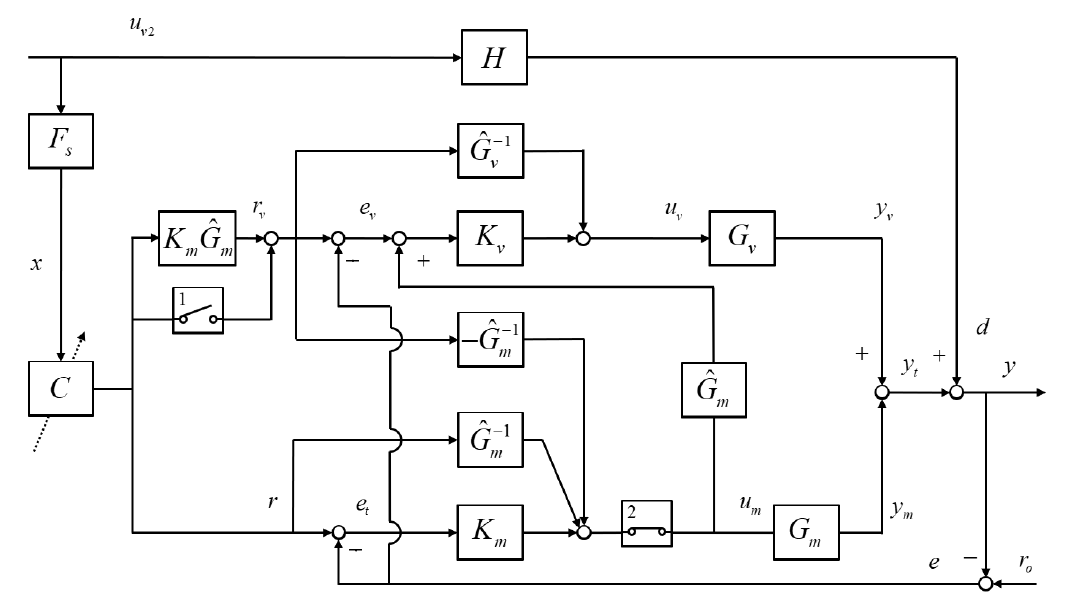}
    \caption{Block diagram of the adaptive feedforward control system with two switches.}
    \label{fig:Pretraining}
\end{figure}
The block diagram of the adaptive feedforward control system with a single-stage/dual-stage
switching fashion is shown in fig.\ref{fig:Pretraining}. Comparing to the block diagram in fig.\ref{fig:AdaptiveControlStructure}, two switches are added. Switch 2 is used to turn on/off the MA, and switch 1 is used to adjust the transfer function of the secondary path such that the control parameters converge to the same stationary point both stages.

\begin{figure}[thpb]
    \centering
    \includegraphics[width=\columnwidth]{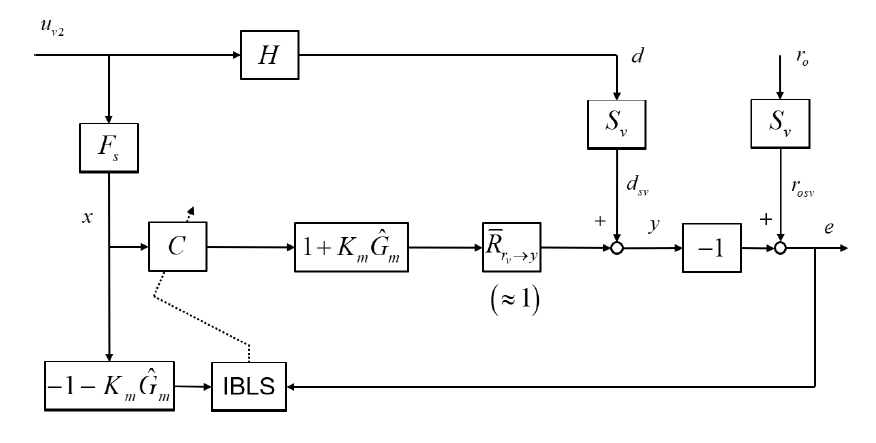}
    \caption{Simplified block Diagram of the adaptive feedforward control system working in
the single-stage mode.}
    \label{fig:SimplifiedBDofACSSingleStage}
\end{figure}

When the Switch 1 is on and Switch 2 is off, the system is running in the pretraining
stage with a single-stage actuator. A simplifed block diagram is shown in fig.\ref{fig:SimplifiedBDofACSSingleStage}. In fig.\ref{fig:SimplifiedBDofACSSingleStage} $S_v$ denotes the closed-loop sensitivity function of the single stage actuator, which
is given by
\begin{equation}
    S_v = \frac{1}{1+G_vK_v}
\end{equation}
$\Bar{R}_{r_v \rightarrow y}$ denotes the transfer function from $r_v$ to $y$, which is given by
\begin{align}
    \Bar{R}_{r_v \rightarrow y} &= S_v(G_v \hat{G}_v^{-1}+G_vK_v) \nonumber \\
    &=\frac{G_v \hat{G}_v^{-1}+G_vK_v}{1+G_vK_v}
\end{align}
If $G_v \hat{G}_v^{-1} \approx 1$, then $\Bar{R}_{r_v \rightarrow y} \approx 1$.
The overall transfer function from $u_{v2}$ to $y$ in a single-stage actuator system is given by
\begin{align}
    \Bar{R}_{u_{v2} \rightarrow y} &= S_v H+ \Bar{R}_{r_v \rightarrow y}(1+K_m\hat{G}_m)CF_s \nonumber \\ &\approx \frac{H}{1+G_vK_v}+(1+K_m\hat{G}_m)CF_s
\end{align}
Recall that the overall transfer function from $u_{v2}$ to $y$ in the dual-stage actuator system
shown in fig.\ref{fig:SimplifiedBDofACS} can be obtained as follows
\begin{align}
    R_{u_{v2} \rightarrow y} &= SH+(R_{r_v \rightarrow y}K_m \hat{G}_m+R_{r \rightarrow y})CF_s \nonumber \\
    &\approx \frac{H}{1+G_vK_v+G_mK_m+G_vK_v\hat{G}_mK_m}+CF_s \nonumber \\
    &\approx \frac{1}{1+K_m\hat{G}_m}\left[\frac{H}{1+G_vK_v}+(1+K_m\hat{G}_m)CF_s\right] \nonumber \\
    &\approx \frac{1}{1+K_m\hat{G}_m}\Bar{R}_{u_{v2} \rightarrow y}
\end{align}
If there exists a controller $C$ that perfectly cancels the imparted disturbance for a single-stage
actuator system and make the transfer function $\Bar{R}_{u_{v2} \rightarrow y}$ equal to zero, then the transfer function $R_{u_{v2} \rightarrow y}$ is also close to zero. In other words, the optimal control for the single-stage actuator system is close to the optimal controller for the dual-stage actuator system. Therefore, the pretrained controller parameters would be good initial points for the finetuning stage, and the output of the MA is kept within its stroke limits during the entire parameter adaptation process.


\section{Results}\label{sec:Results}
\begin{figure}[H]
    \centering
    \includegraphics[width=0.9\columnwidth]{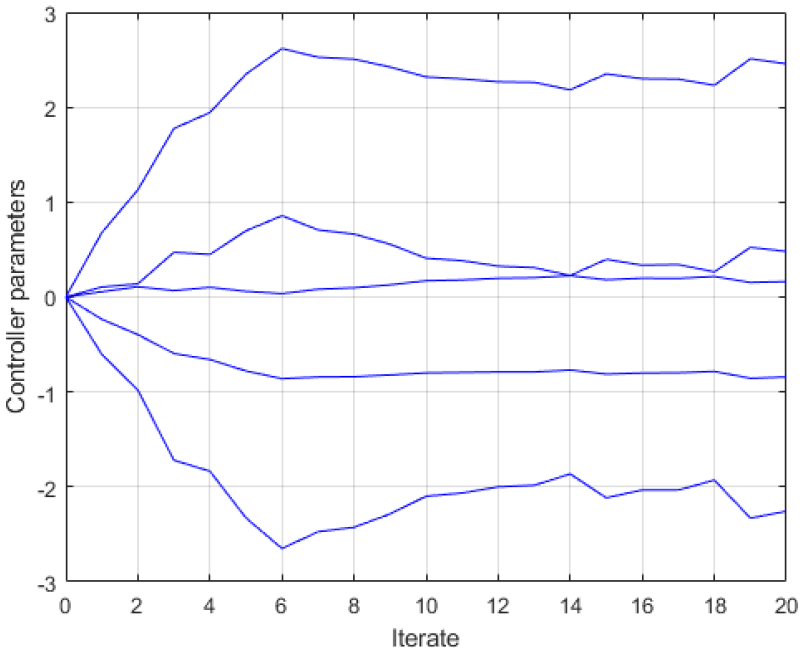}
    \caption{Convergence of controller parameters in the single-stage mode.}
    \label{fig:ParameterConvergence}
\end{figure}

\begin{figure}[H]
    \centering
    \includegraphics[width=0.9\columnwidth]{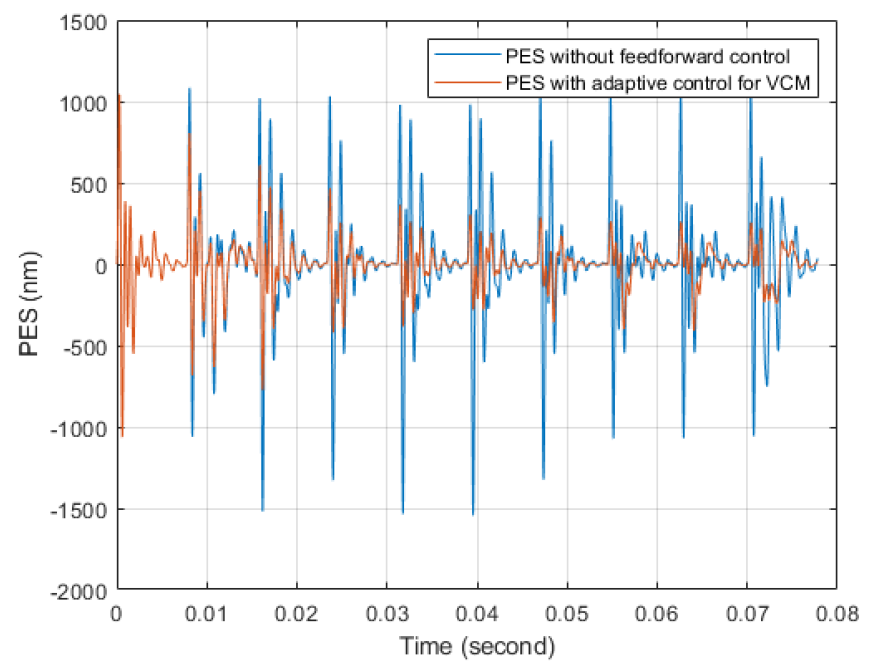}
    \caption{Position error signals in single-stage mode.}
    \label{fig:PESSingleStage}
\end{figure}

\begin{figure}[H]
    \centering
    \includegraphics[width=0.9\columnwidth]{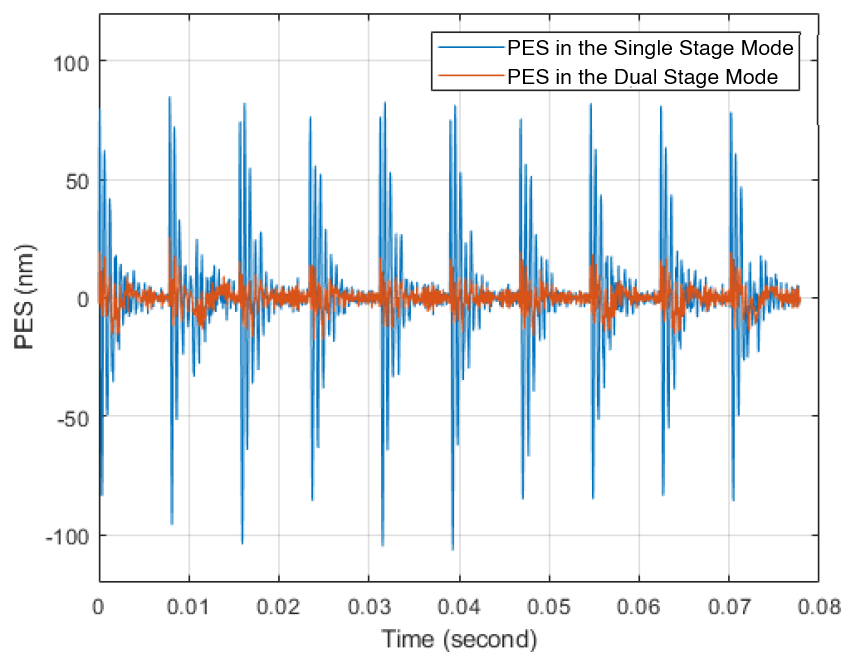}
    \caption{Position error signals in dual-stage mode.}
    \label{fig:PESDualStage}
\end{figure}

The adaptive feedforward reference design presented was numerically implemented with realistic plant models for VCM, MA, disturbance cross transfer function $H$ (shown in appendix) and $4^{th}$ order FIR filter for the adaptive controller. Fig. \ref{fig:ParameterConvergence} shows the controller parameters in each iteration. 
Fig.\ref{fig:PESSingleStage} shows the PES during the single-stage mode i.e., the pretraining stage. As only the VCM is turned on in this case, compensation is not very accurate, but the controller parameters converge to the optimal values. The converged parameters were used to initialize the controller parameters in the adaptation process for the dual-stage mode. Fig.\ref{fig:PESDualStage} shows the PES in the dual-stage mode which is within the desired limits.

\section{Conclusions}\label{sec:Conc}
In this paper an adaptive feedforward control structure has been developed to learn and reject the disturbance process emanating from the dual stage actuator in the track seeking mode and affecting the dual stage actuator in the track following mode in a multi actuator hard disk drive. This control structure is designed as an add on controller to the already existing feedback control structure. The controller is designed as an FIR filter with unknown coefficients that are adapted to minimize the position error signal. A feedforward reference design structure has been used to make the adaptive controller robust to plant variations and to keep the output of the micro actuator within its stroke limits.
\section*{ACKNOWLEDGMENT}
We would like to acknowledge the feedback and financial support from ASRC hosted by the International Disk Drive Equipment and Materials Association (IDEMA).


\section*{Appendix}\label{sec:Appendix}
\begin{figure}[htpb]
    \centering
    \includegraphics[width=0.9\columnwidth]{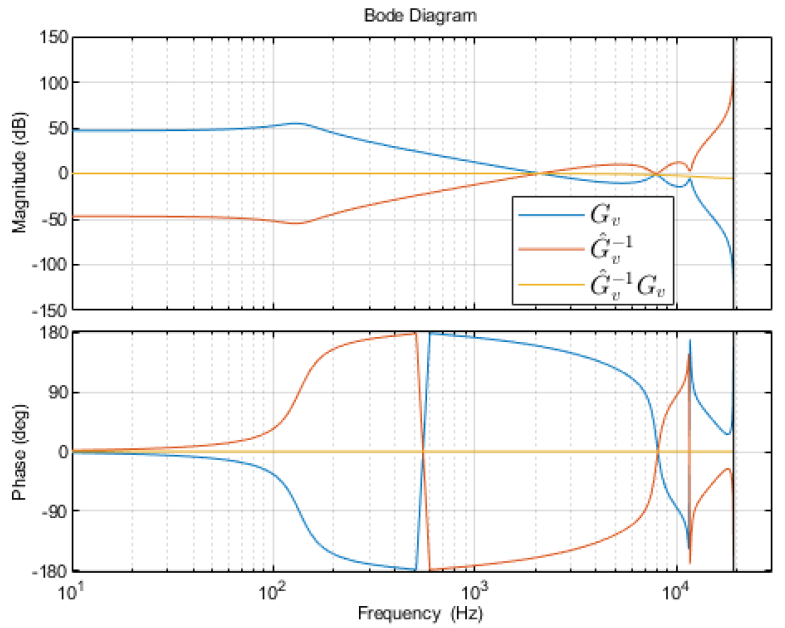}
    \caption{ Frequency responses for $G_v$,$\hat{G}_v^{-1}$ and $\hat{G}_v^{-1}G_v$}
    \label{fig:VCM}
\end{figure}
\begin{figure}[H]
    \centering
    \includegraphics[width=0.9\columnwidth]{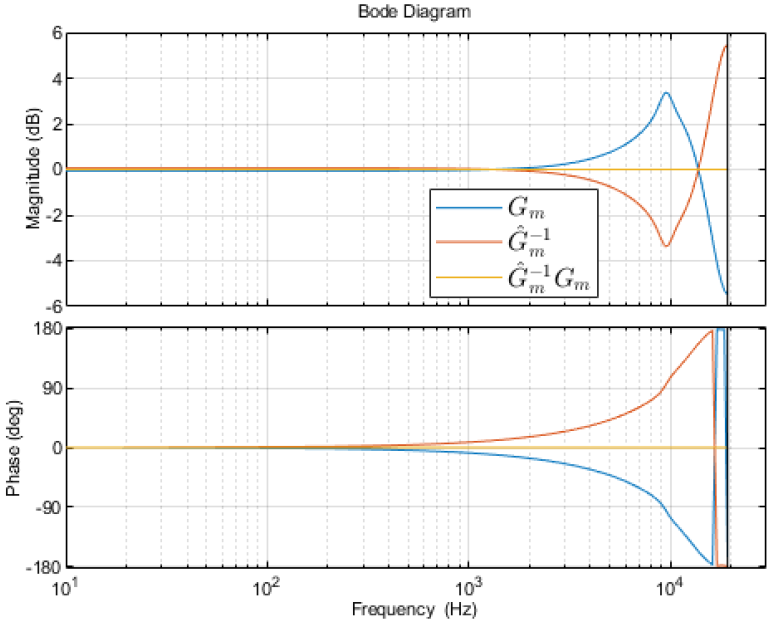}
    \caption{ Frequency responses for $G_m$,$\hat{G}_m^{-1}$ and $\hat{G}_m^{-1}G_m$}
    \label{fig:MA}
\end{figure}
\begin{figure}[H]
    \centering
    \includegraphics[width=0.9\columnwidth]{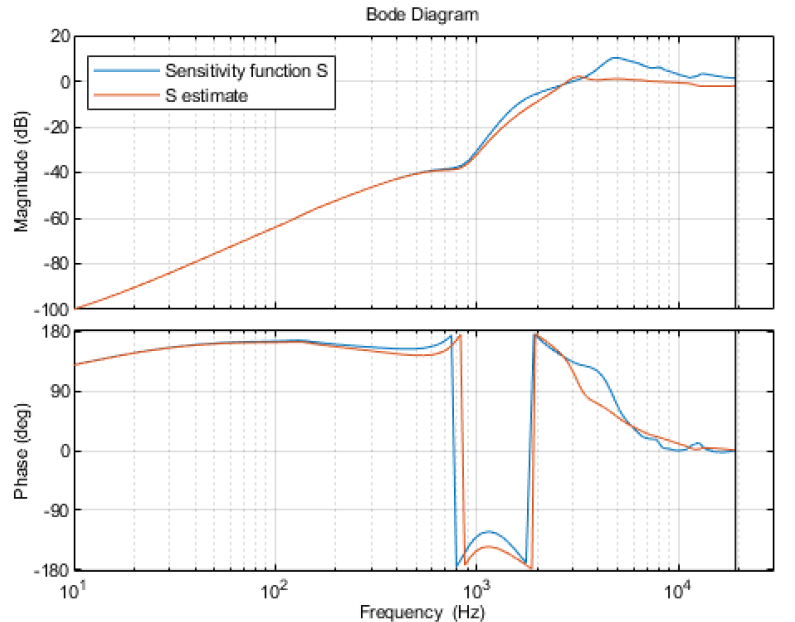}
    \caption{ Frequency responses for $S$,$\hat{S}$}
    \label{fig:Sensitivity}
\end{figure}
\begin{figure}[H]
    \centering
    \includegraphics[width=0.9\columnwidth]{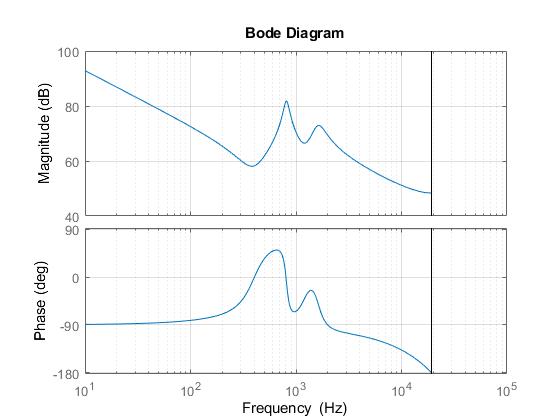}
    \caption{ Frequency response of the cross transfer function H}
    \label{fig:H}
\end{figure}

\begin{figure}[H]
    \centering
    \includegraphics[width=0.9\columnwidth]{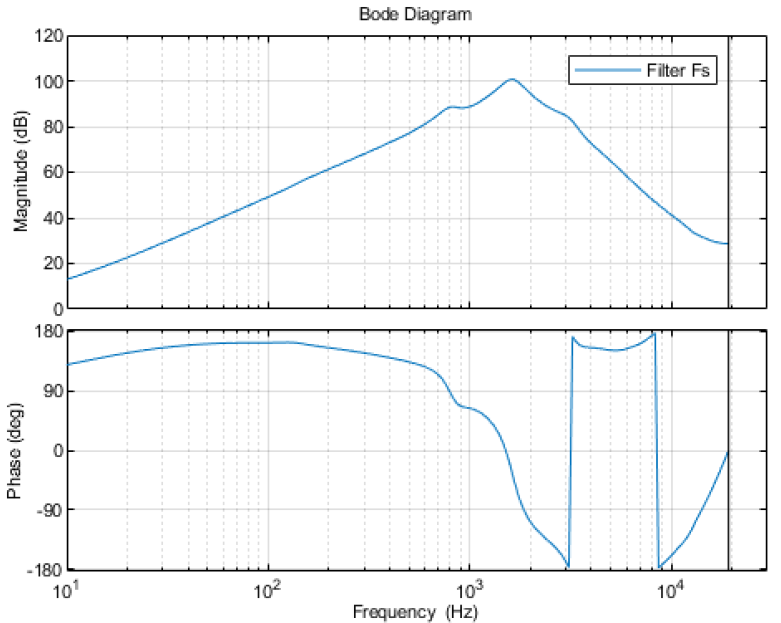}
    \caption{Frequency response of filter $F_s$}
    \label{fig:Fs}
\end{figure}

\end{document}